\documentclass[hyper]{JHEP} 

\usepackage{epsfig}

\newcommand\fverb{\setbox\pippobox=\hbox\bgroup\verb}

\newcommand\fverbdo{\egroup\medskip\noindent%

            \fbox{\unhbox\pippobox}\ }

\newcommand\fverbit{\egroup\item[\fbox{\unhbox\pippobox}]}

\newbox\pippobox

\title{New Models of
 $f(R)$ Theories of Gravity}
\author{J. Kluso\v{n}\\
Department of
Theoretical Physics and Astrophysics\\
Faculty of Science, Masaryk University\\
Kotl\'{a}\v{r}sk\'{a} 2, 611 37, Brno\\
Czech Republic\\
E-mail: \email{klu@physics.muni.cz}}
\preprint{\hepth{0910.5852}}

 \abstract{We introduce
 new models of $f(R)$ theories of gravity that
 are generalization of Ho\v{r}ava-Lifshitz gravity.
  } \keywords{Ho\v{r}ava-Lifshitz gravity}

\def\mH{\mathcal{H}}

\def\bx{\mathbf{x}}
\def\by{\mathbf{y}}

\newcommand{\mG}{\mathcal{G}}

\newcommand{\bT}{\mathbf{T}}

\newcommand{\mL}{\mathcal{L}}

\def\pb #1{\left\{#1\right\}}

\begin{document}
\section{Introduction and Summary}\label{first}
The acceleration of the universe was
discovered 10 years ago and its
explanation is one of the most secret
enigmas of current theoretical physics.
According to \cite{Riess:1998cb} and
the other experiments  that map the
anisotropies of the cosmic microwave
background, the approximately $76 \% $
of the energy content of the universe
is not dark or luminous matter but it
is instead a mysterious form of dark
energy that is exotic, invisible and
unclustered. In order to explain the
origin of this form of matter three
main classes of models for this
acceleration exist:
\begin{enumerate}
\item A cosmological constant $\Lambda$
\item Dark Energy
\item Modified Gravity
\end{enumerate}
The cosmological constant is the most
obvious explanation of the acceleration
of the universe
\cite{Padmanabhan:2002ji,Carroll:2000fy}.
However $\Lambda$ suffers from well
known cosmological problem that
requires extreme fine-tuning.

The second class of models postulates
the existence of a dark energy fluid
with equation of state $P\approx -\rho$
where $\rho$ and $P$ are the energy
density and pressure of the fluid
\cite{Copeland:2006wr}.

The last class of models known as
Extended Theories
of Gravity (ETG) corresponds to
the modification of the action of the
gravitational fields.  These theories are
based on idea of extension of the
Einstein Hilbert action by adding
higher order curvature invariants
 such as
$R^2,R^{\mu\nu}R_{\mu\nu}\dots$. and/or
minimally or non-minimally coupled
scalar fields to the dynamics
\footnote{For review and extensive list
of references, see
\cite{Capozziello:2009nq,Sotiriou:2008rp,Nojiri:2008nt,Faraoni:2008mf,Nojiri:2006ri}.}.
In this paper we consider the models
corresponding to $f(R)$ theories that
are general functions of  the Riemann
scalar $R$.

As we said above $f(R)$ theories should
mainly explain the mysterious
acceleration of the Universe and hence
provide modification of the Einstein
theory on large scales. On the other
hand they do not solve the second
fundamental problem considering theory
of gravity  which is the fact that
the quantum gravity is
 non-renormalizable theory. In order to
improve this situation new formulation
of the quantum gravity theory was
proposed recently in series of papers
\cite{Horava:2009uw,Horava:2008ih,Horava:2009if,Horava:2008jf}
\footnote{For detailed study of this
proposal, see
\cite{Koyama:2009hc,Park:2009gf,Chen:2009vu,Wang:2009azb,Suyama:2009vy,
Tang:2009bu,Boehmer:2009yz,Leon:2009rc,Blas:2009qj,Wu:2009ah,Iorio:2009qx,
Ding:2009pq,Carloni:2009jc,Lee:2009rm,Kluson:2009rk,Wang:2009yz,Peng:2009uh,
Mukohyama:2009tp,Blas:2009yd,Germani:2009yt,Myung:2009sa,Park:2009zra,
Sakamoto:2009ww,Nojiri:2009th,Calcagni:2009qw,Mukohyama:2009mz,Gao:2009ht,
Kim:2009zn,Sotiriou:2009bx,Li:2009bg,Cai:2009dx,Mukohyama:2009zs,Sotiriou:2009gy,
Gao:2009bx,Piao:2009ax,Cai:2009ar,Nastase:2009nk,Brandenberger:2009yt,
Mukohyama:2009gg,Lu:2009em,Kluson:2009sm,Kiritsis:2009sh,Calcagni:2009ar,
Takahashi:2009wc,Visser:2009fg,Kocharyan:2009te,Kiritsis:2009rx,
Chen:2009jr,Saridakis:2009bv,Bogdanos:2009uj,Cai:2009in,Chakrabarti:2009ku}.}.
The key idea of this proposal is to
abandon the fundamental role of the
local Lorentz invariance and assume
instead that this appears only at low
energies as an approximative symmetry.
The breaking of Lorentz invariance is
achieved by introducing additional
structure to the space-time that is a
preferred foliation by $3-$dimensional
space-like surfaces and then splitting
of the coordinates into space and time.
This procedure allows to complete the
Einstein's general relativity with
higher spatial derivatives of the
metric and then improve the UV behavior
of the graviton propagator and makes
the theory renormalizable by
power-counting.

Even if the Ho\v{r}ava-Lifshitz theory
has many attractive properties
investigations of various aspects of
Ho\v{r}ava gravity have started to
reveal some potentially troublesome
features. The first one is considering
the \emph{detailed balance condition}
that strongly restricts the form of the
potential terms in the gravity action.
However the detailed investigation of
the Ho\v{r}ava-Lifshitz theory
performed in past few months strongly
suggests  that this theory has much
better behavior when we abandon the
condition of the detailed balance
\footnote{For nice discussion, see
\cite{Sotiriou:2009bx}.}.

The second fundamental ingredient of
the original Ho\v{r}ava-Lifshitz was
the \emph{projectability condition}
that says the lapse function in ADM
formulation of gravity depends on time
only. Even if for the most important
vacuum and cosmological solutions
Einstein equations can be put into the
form of constant lapse function it
seems that the consistency of the
Ho\v{r}ava-Lifshitz theory requires the
projectability condition. In fact, the
investigation of the algebra of
constraints of the Ho\v{r}ava-Lifshitz
theory  without projectability
condition performed in  paper
\cite{Li:2009bg} implies some
pathological behavior of this theory.
 On the other hand it was already shown in
\cite{Horava:2008ih} that
   the algebra of constraints of
Ho\v{r}ava-Lifshitz theory with
projectability condition is closed.
However it is important to  stress that
we should be very careful considering
the definite conclusion about the
consistency/non-consistency of
Ho\v{r}ava-Lifshitz theory with/without
projectability condition as was
discussed recently in
\cite{Park:2009gf,Blas:2009yd} in
relation with an another  issue of
Ho\v{r}ava-Lifshitz theory that is
spin-0 scalar graviton with addition to
the standard spin-2 tensor graviton.
The study of the properties of this
mode showed that it persists down to
low energies and exhibits the
pathological behavior that should be
disaster for the consistency of
Ho\v{r}ava-Lifshitz theory. Explicitly,
it was shown in \cite{Blas:2009yd} that
this mode is strongly coupled at
energies above a very low energy scale
and it suffers from fast instabilities.
It turns out that these properties
appear both in the non-projectable and
projectable cases. On the other hand it
was argued in very recent paper
\cite{Blas:2009qj} that it is possible
to extend the original
Ho\v{r}ava-Lifshitz action in such a
way that these dangerous properties are
absent.

Despite of these open problems it is
clear that Ho\v{r}ava-Lifshitz theory
is very interesting  and stimulating
 proposal that
should be extended in many directions.
Our goal is to generalize this theory
in the similar way as the ordinary
Einstein gravity  is extended   to
$f(R)$ theories. We started this
program in our previous paper
\cite{Kluson:2009rk} where we formulated
the Hamiltonian form of  $f(R)$
Ho\v{r}ava-Lifshitz theory based on the
condition of the detailed balance. We
considered one explicit example of
$f(R)$ Ho\v{r}ava-Lifshitz theory with
the function $f(x)=\sqrt{x}$ and
 we   found the
Lagrangian version of this theory. In
this paper we generalize this analysis
to the case of arbitrary $f(R)$
Hamiltonian form of Ho\v{r}ava-Lifshitz
theory. We show that when we extend the
original Hamiltonian of $f(R)$
Ho\v{r}ava-Lifshitz theory by
introducing two non-propagating degrees
of freedom we can find the Hamiltonian
in such a form that allows us to
perform the Legendre transformation
from the Hamiltonian to the
 Lagrangian description. Then solving
the equations of motion for these
auxiliary fields we finally find a new
form of $f(R)$ theories that are
invariant under foliation of preserving
diffeomorphism \cite{Horava:2008ih}.

It is also possible to introduce the
Ho\v{r}ava-Lifshitz theories of gravity
in an alternative way when we replace
4-dimensional Ricci scalar $R$ with
combination $R\rightarrow
K_{ij}\mG^{ijkl}K_{kl}+ \dots$ known
from the linear Ho\v{r}ava-Lifshitz
theory of gravity. The resulting theory
is invariant under foliation preserving
diffeomorphism.

We hope that this short note could be
the starting point
in the study of the new form of $f(R)$
theories of gravity. It will be
certainly very interesting to study the
equations of motion that follow from these
actions
and find their possible cosmological
solutions. It would be also interesting
to perform the analysis of the fluctuation
modes around  classical background
and study their physical properties.
 We hope to return to these
 problems in future.
\section{The First Version of $f(R)$
Ho\v{r}ava-Lifshitz Theory of Gravity}
In this section we consider
 $f(R)$ Ho\v{r}ava-Lifshitz
theory of gravity that was introduced
in \cite{Kluson:2009rk}. These theories
are defined by Hamiltonian
\begin{equation}\label{defH}
H=\int d^D\bx \left(N(t)
\mH_0(t,\bx)+N_i(t,\bx) \mH^i(t,\bx)\right)
\ ,
\end{equation}
where $\mH^i$ is the generator of the
spatial diffeomorphism
\begin{equation}
\mH^i(t,\bx)=-2\nabla_j\pi^{ji}(\bx) \
\end{equation}
that takes the same form as in standard
Einstein theory of gravity. On the
other hand  the Hamiltonian constraint $\mH_0$
of  $f(R)$ Ho\v{r}ava-Lifshitz
 $\mH_0$ is defined as
\begin{equation}\label{defmH0}
\mH_0(t,\bx)=
\kappa^2 \sqrt{g}f\left( Q^{\dag ij}
\frac{1}{g}\mG_{ijkl} Q^{kl}\right)
 \ ,
\end{equation}
where $\kappa$ is the coupling constant
of the theory, $g=\det g_{ij}$  and
where $f$ is an arbitrary function.
  Further, the functions
 $Q^{ij}$ and
$Q^{\dag ij}$ are defined as
\begin{equation}
Q^{ ij}=i\pi^{ij}+ \sqrt{g}E^{ij} \ ,
\quad Q^{\dag ij}=-i\pi^{ij}+
\sqrt{g}E^{ij} \ , \quad
\end{equation}
where $g_{ij},\ , i,j=1,\dots,D$ are
components of metric and  $\pi^{ij}$
are conjugate momenta. These canonical
variables have non-zero   Poisson
brackets
\begin{equation}
\pb{g_{ij}(\bx),\pi^{kl}(\by)}=
\frac{1}{2}(\delta_i^k\delta_j^l+
\delta_i^l\delta_j^k)\delta(\bx-\by) \ .
\end{equation}
Finally we  explain the meaning of the
objects $E^{ij}$.  In the original
version of
 Ho\v{r}ava-Lifshitz theory
 based on the \emph{the condition
 of the detailed balance}
they are given by variation of the
$D$-dimensional functional $W =\int
d^D\bx \sqrt{g}w(g,\partial g)$
\begin{equation}
\sqrt{g}E^{ij}(\bx)=\frac{1}{2}\frac{
\delta W}{\delta g_{ij}(\bx)} \  .
\end{equation}
 However we can clearly
abandon this requirement and consider
the generalized Ho\v{r}ava-Lifshitz
theories where the Hamiltonian
constraint $\mH_0$ takes the form
\begin{equation}
\mH_0=\kappa^2 \sqrt{g}
f\left(\frac{1}{g}\pi^{ij}\mG_{ijkl}\pi^{kl}+
E^{ij}\mG_{ijkl}E^{kl}+\dots\right) \ ,
\end{equation}
where $\dots$ denote the contribution
from additional terms that violet the
detailed balance condition and that are
crucial for reproducing General
relativity at low energy regime.
 However for simplicity of
the notation we consider the first form
of the Hamiltonian (\ref{defmH0})
keeping in mind that the analysis below
is valid for the general case as well.
Finally the  "metric on the space of
metric", $\mathcal{G}^{ijkl}$ is
defined as
\begin{equation}\label{DeW}
\mG^{ijkl}=\frac{1}{2}(g^{ik}g^{jl}+g^{il}
g^{jk}-\lambda g^{ij}g^{kl}) \
\end{equation}
with $\lambda$ an arbitrary real
constant. The inverse $\mG_{ijkl}$
takes the form
\begin{equation}\label{DeWi}
\mG_{ijkl}=\frac{1}{2}(g_{ik}g_{jl}+
g_{il}g_{jk})-\tilde{\lambda}g_{ij}g_{kl} \ ,
\end{equation}
where $\tilde{\lambda}=\frac{\lambda}{D\lambda-1}$.
 Note that (\ref{DeW})
together with (\ref{DeWi}) obey the
relation \footnote{We use the
terminology introduced in
\cite{Horava:2008ih} and that we review
there.
 In case of relativistic
theory, the full diffeomorphism
invariance fixes the value of $\lambda$
uniquely to equal $\lambda=1$. In this
case the object $\mG_{ijkl}$ is known
as the "De Witt metric". We use this
terminology to more general case when
$\lambda$ is not necessarily equal to
one.}
\begin{equation}
\mG_{ijmn}\mG^{mnkl}=\frac{1}{2}
(\delta_i^k\delta_j^l+\delta_i^l\delta_j^k)
\ .
\end{equation}
We should stress one important point.
The Hamiltonian (\ref{defH}) contains
the lapse function that depends on $t$
only. This  property is known as
\emph{projectability condition}. It is
still an open problem whether this
condition should be imposed or whether
we should consider the lapse function
with the full space and time
dependence. In this section we consider
the theories that obey
\emph{projectability condition}. When
this condition is imposed
 the
functions $N(t)$ and $N_i(t,\bx)$ have
conjugate momenta $\pi_N, \pi^i(\bx)$ that
form the primary constraints of the theory
\begin{equation}
\pi_N\approx 0 \ , \quad
\pi^i(\bx)\approx 0 \ .
\end{equation}
Since these constraints
 have to be preserved during the
time evolution of the system  we find
\begin{equation}\label{bTT}
\partial_t \pi_N=
\pb{\pi_N,H}=\int d^D\bx \mH_0(\bx)\equiv
\bT_T \approx 0 \ .
\end{equation}
Note that the global constraint $\bT_T\approx 0$
is fundamentally different from the
local constraint $\mH_0(\bx)\approx 0$
that is imposed in case when the theory
does not respect the projectability
condition.  Further, the consistency of
 the constraints
$\pi^i(\bx)\approx 0$ imply the
secondary constraints
\begin{equation}
\partial_t \pi^i(\bx)=
\pb{\pi^i(\bx),H}=
-\mH^i(\bx)\approx 0 \ .
\end{equation}
It is natural to introduce the smeared
form of the spatial diffeomorphism constraint
\begin{equation}\label{bTS}
\bT_S\equiv \int d^D\bx
\zeta_i(\bx)\mH^i(\bx) \ .
\end{equation}
The next goal is to calculate the
Poisson bracket of constraints
(\ref{bTT}) and (\ref{bTS}). These
brackets have been determined in
\cite{Kluson:2009rk} with following
result
\begin{eqnarray}
\pb{\bT_T,\bT_T}&=&0 \ , \nonumber \\
\pb{\bT_S(\zeta),\bT_T}&=&0 \ ,
\nonumber
\\
\pb{\bT_S(\zeta),\bT_S(\xi)}&=&
\bT_S(\zeta^i\partial_i\xi^k-\xi^i\partial_i\zeta^k)
\ .
\nonumber \\
\end{eqnarray}
In other words the constraint algebra
of the gravity action that is invariant
under foliation preserving
diffeomorphism is closed. This is
a crucial simplification with respect
to theories that do not respect the
projectability condition. More
precisely, the algebra of constraints
of the Ho\v{r}ava-Lifshitz theory
without projectability condition was
 calculated in \cite{Li:2009bg} with
 the result that suggests inconsistency
 of this theory.
\subsection{Lagrangian
Formulation}\label{third}
 In this section we formulate the Lagrangian
formalism for the $f(R)$ theories
defined by the Hamiltonian
(\ref{defH}). In order to this we argue
that the Hamiltonian constraint $\mH_0$
defined in (\ref{defH}) can be written
as
\begin{eqnarray}\label{defH0A}
\mH_0=\kappa^2\sqrt{g}\left[B\left(
Q^{\dag ij}\frac{1}{g}\mG_{ijkl}
Q^{kl}-A\right)+f(A)\right]+v_A P_A+v_B
P_B \ ,
\nonumber \\
\end{eqnarray}
where we introduced two new fields
$A(\bx),B(\bx)$ with conjugate momenta
$P_A(\bx),P_B(\bx)$ and corresponding
Poisson brackets
\begin{equation}
\pb{A(\bx),P_A(\by)}=\delta(\bx-\by) \
, \quad
\pb{B(\bx),P_B(\by)}=\delta(\bx-\by) \
.
\end{equation}
Looking on the form of the Hamiltonian
density (\ref{defH0A}) we see that it
is natural to  interpret
 $P_A(\bx),P_B(\bx)$ as the
primary constraints of the theory
\begin{equation}\label{PAcon}
P_A(\bx)\approx 0 \ , \quad
P_B(\bx)\approx 0 \ .
\end{equation}
Then the requirement  that the
constraints (\ref{PAcon}) are preserved
during the time evolution implies an
existence of the secondary constraints
$G_A(\bx),G_B(\bx)$ since
\begin{eqnarray}
\partial_t P_A&=&\pb{P_A,H}=
\kappa^2\sqrt{g}(B-f'(A))\approx
\kappa^2\sqrt{g} G_A\approx 0 \ ,
\nonumber \\
\partial_t P_B&=&\pb{P_B,H}=
-\kappa^2\sqrt{g}(Q^{\dag
ij}\frac{1}{g} \mG_{ijkl}
Q^{kl}-A)=-\kappa^2\sqrt{g}G_B\approx 0 \ ,  \nonumber \\
\end{eqnarray}
where $H$ is given in (\ref{defH}) with
$\mH_0$ defined in (\ref{defH0A}). Let
us now calculate the Poisson brackets
between $P_A,P_B$ and $G_A,G_B$. It is
a straightforward exercise to calculate
them and we find
\begin{eqnarray}\label{defPBsecond}
\pb{P_A(\bx),G_B(\by)}&=&\delta(\bx-\by)
\ ,
\nonumber \\
\pb{P_A(\bx),G_A(\by)}&=&f''(A)\delta(\bx-\by)
\ ,
\nonumber \\
\pb{P_B(\bx),G_A(\by)}&=&-\delta(\bx-\by)
\ ,
\nonumber \\
\pb{P_B(\bx),G_B(\by)}&=&0 \ .
\nonumber \\
\end{eqnarray}
These results imply that
$P_A,P_B,G_A,G_B$ form the second class
constraints. As a consequence the
requirement of the preservation of the
constraints $G_A,G_B$ under the time
evolution fixes $v_A,v_B$ uniquely.
Since  these constraints are second
class they strongly vanish and can be
explicitly solved with the result
\begin{equation}
B=f'(A) \ , \quad  A=Q^{ij
\dag}\frac{1}{g} \mG_{ijkl} Q^{kl} \ .
\end{equation}
Then if we insert these results into
the Hamiltonian density given in
(\ref{defH0A}) we find that it takes
the form
\begin{equation}
\mH_0=\kappa^2\sqrt{g} f\left(Q^{\dag
ij}\frac{1}{g}\mG_{ijkl}
Q^{kl}\right) \nonumber \\
\end{equation}
that coincides with (\ref{defH}). As
the final point note that for the
system with the second class
constraints we have to introduce the
Dirac brackets instead of Poisson
brackets
\begin{eqnarray}
\pb{g_{ij}(\bx),\pi^{kl}(\by)}_D&=&
\pb{g_{ij}(\bx),\pi^{kl}(\by)}-
\nonumber \\
&-&\int d\bx' d\by' \pb{g_{ij}(\bx),
\Phi_I(\bx')}\Omega^{IJ}(\bx',\by')
\pb{\Phi_J(\by'),\pi^{kl}(\by)} \ , \
\nonumber \\
\end{eqnarray}
where $\Phi_I, I=1,2,3,4$ is a
collection of the second class
constraints $\Phi_I= (P_A,G_A,P_B,G_B)$
and where $\Omega^{IJ}$ is inverse of
the matrix of Poisson brackets
(\ref{defPBsecond}). Explicitly
\begin{equation}
\Omega_{IJ}(\bx,\by)\equiv
\pb{\Phi_I(\bx),\Phi_J(\by)}=\left(\begin{array}{cccc}
0
& f'' & 0 & 1 \\
-f'' & 0 & 1 & 0 \\
0 & -1 & 0 & 0 \\
-1 & 0 & 0 & 0 \\
\end{array}\right)\delta(\bx-\by)
\end{equation}
so that
\begin{equation}
\Omega^{IJ}(\bx,\by)=
\left(\begin{array}{cccc} 0 & 0 & 0 & 1
\\
0 & 0 & - 1 & 0 \\
0 & 1 & 0 & -f'' \\
1 & 0 & f'' & 0 \\
\end{array}\right)\delta(\bx-\by)
\end{equation}
However due to the structure of the
matrix $\Omega^{IJ}$ and  the fact that
$g_{ij}$ and $\pi^{ij}$ have vanishing
Poisson brackets with $P_A,P_B$ we find
that the Dirac brackets coincide with
Poisson brackets.

In summary, we claim that  the
 Hamiltonian density that is useful for
 the developing of the Lagrangian
 formalism takes the form
\begin{equation}\label{mHa}
\mH=-2N_i\nabla_j
\pi^{ij}+N\kappa^2\sqrt{g}\left[B\left(
Q^{\dag ij}\frac{1}{g}\mG_{ijkl}
Q^{kl}-A\right)+f(A)\right]+v_A P_A+v_B
P_B \ .
 \end{equation}
The main advantage of the Hamiltonian
(\ref{mHa}) density with respect to
(\ref{defH}) is that it
 depends on  the momenta $\pi^{ij}$
 quadratically and hence the Legendre
transformation from the Hamiltonian to
the Lagrangian description can be
easily performed.
 Explicitly, the
canonical equations of motion for $A,B$
and $g_{ij}$ take the form
\begin{eqnarray}\label{partAB}
\partial_t A&=&\pb{A,H}=
v_A  \ , \quad
\partial_t B=\pb{B,H}=v_B \ ,
\nonumber \\
\partial_t g_{ij}&=&\pb{g_{ij},H}=
\frac{2BN\kappa^2}{\sqrt{g}}\mG_{ijkl}\pi^{kl}+
\nabla_i N_j+\nabla_j N_i \ .
\end{eqnarray}
It is natural to define
\begin{equation}
K_{ij}=\frac{1}{2N} (\partial_t
g_{ij}-\nabla_i N_j-\nabla_j N_i)
\end{equation}
so that  the equation on the second
line in (\ref{partAB}) implies
\begin{eqnarray}
 K_{ij}=
\frac{B\kappa^2}{\sqrt{g}}\mG_{ijkl}\pi^{kl}
\ .
\nonumber \\
\end{eqnarray}
Then it is easy to determine the
corresponding Lagrangian
\begin{eqnarray}\label{LFR}
L&=&\int  d^D\bx(\partial_t
g^{ij}\pi_{ij}+\partial_t AP_A+
\partial_t BP_B-\mH)=\nonumber \\
&=&\int d^D\bx
N\sqrt{g}\left(\frac{1}{B\kappa^2}K_{ij}\mG^{ijkl}
K_{kl}- \kappa^2B(
E^{ij}\mG_{ijkl}E^{kl}-
A)-\kappa^2f(A)\right) \ .
\nonumber \\
\end{eqnarray}
Finally we eliminate non-dynamical
fields $A$ and $B$ from (\ref{LFR}).
Varying  (\ref{LFR}) with respect to
$B$
\begin{eqnarray}\label{eqB}
-\frac{1}{B^2\kappa^2} K_{ij}\mG^{ijkl}
K_{kl}-( E^{ij}\mG_{ijkl}E^{kl}- A)=0 \
,
\nonumber \\
\end{eqnarray}
and varying (\ref{LFR}) with respect to
$A$ we find
\begin{equation}\label{eqA}
 B-f'(A)=0 \ .
 \end{equation}
Inserting (\ref{eqA}) into (\ref{eqB})
we obtain the equation
\begin{eqnarray}
-\frac{1}{f'^2(A)\kappa^2}K_{ij}\mG^{ijkl}
K_{kl}-\kappa^2(E^{ij}\mG_{ijkl}E^{kl}-A)=0
\nonumber \\
\end{eqnarray}
that, for known function $f$, allows us
to express $A$ as a function of
$K_{ij},g_{ij}$
\begin{equation}
A=\Psi(K_{ij},g_{ij}) \ .
\end{equation}
Inserting these results into the
Lagrangian (\ref{LFR})  we find the new
class of  $F(R)$ theories of gravity in
the form
\begin{eqnarray}\label{LFRf}
L&=&\int d^D \bx \mL \ , \nonumber \\
\mL&=&
N\sqrt{g}\left[
\frac{2}{f'(\Psi(K_{ij},g_{ij}))\kappa^2}K_{ij}\mG^{ijkl}
K_{kl}-\kappa^2
f(\Psi(K_{ij},g_{ij}))\right] \ .
\nonumber \\
\end{eqnarray}
Let us consider the specific example of
the theory defined by the function
$f(A)=\sqrt{1+A}-1$. For this function
the equation (\ref{eqA}) implies
\begin{equation}
B=\frac{1}{2\sqrt{1+A}} \ .
\end{equation}
Inserting this result into
 (\ref{eqB})  we find
\begin{equation}
A=\Psi(g_{ij},K_{ij})=\frac{E^{ij}\mG_{ijkl}E^{kl}
+\frac{4}{\kappa^4}K_{ij}\mG^{ijkl}K_{kl}}
 {1-\frac{4}{\kappa^4} K_{ij}\mG^{ijkl}K_{kl}} \ .
\end{equation}
and hence the Lagrangian density takes
the form
\begin{eqnarray}\label{LDex}
\mL=-N\sqrt{g}(1-4K_{ij}\mG^{ijkl}K_{kl})
\sqrt{1+\Psi}+\kappa^2\sqrt{g}N= \nonumber \\
=-\kappa^2
N\sqrt{g}\left(\sqrt{1-\frac{4}{\kappa^4}
K_{ij}\mG^{ijkl}K_{kl}}
\sqrt{1+E^{ij}\mG_{ijkl}E^{kl}}-1\right)
\ .
\nonumber \\
\end{eqnarray}
We see that (\ref{LDex}) coincides with
the Lagrangian density studied in
\cite{Kluson:2009rk}.

Finally we show that the action $S=\int
dt L$ where $L$ is given in
(\ref{LDex}) is invariant under the
foliation-preserving diffeomorphism
that by definition
 consists a space-time
dependent spatial diffeomorphisms as
well as time-dependent time
reparameterization. These symmetries
are
 generated by infinitesimal
transformations
\begin{equation}\label{fpd}
\delta x^i\equiv
x'^i-x^i=\zeta^i(t,\bx) \ , \quad
\delta t\equiv t'-t=f(t) \ .
\end{equation}
It was shown in \cite{Horava:2008ih}
that the metric transform under
(\ref{fpd}) as
\begin{eqnarray}\label{trm}
g'_{ij}(t',\bx')&=&g_{ij}(t,\bx)-
g_{il}(t,\bx)\partial_j
\zeta^l-\partial_i
\zeta^k g_{kj}(t,\bx)  \ , \nonumber \\
g'^{ij}(t',\bx')&=& g^{ij}(t,\bx)+
\partial_n \zeta^i g^{nj}(t,\bx)
+g^{in}(t,\bx)
\partial_n \zeta^j
 \  \nonumber \\
\end{eqnarray}
and the fields
 $N_i(t,\bx),N(t)$
 transform under (\ref{fpd}) as
\begin{eqnarray}
N'_i(t',\bx')&=& N_i(t,\bx) -N_i(t,\bx)
\dot{f}-N_j(t,\bx)\partial_i
\zeta^j-g_{ij}(t,\bx)
\dot{\zeta}^j \ ,      \nonumber \\
 N'^i(t',\bx')
&=&N^i(t,\bx)+N^j(t,\bx)\partial_j
\zeta^i(t,\bx)-
N^i(t,\bx)\dot{f}-\dot{\zeta}^i(t,\bx)
\ ,
\nonumber \\
N'(t')&=&-N(t) \dot{f} \ . \nonumber \\
\end{eqnarray}
Further, the transformation
property of the metric components (\ref{trm})
imply following transformation
prescription for $\mG_{ijkl}$
\begin{eqnarray}
\mG'_{ijkl}(t',\bx')&=&
\mG_{ijkl}(t,\bx)+\partial_i\zeta^m(t,\bx)
\mG_{mjkl}(t,\bx)
+\partial_j \zeta^m(t,\bx)\mG_{imkl}(t,\bx)+
\nonumber \\
&+&
\mG_{ijml}(t,\bx)
\partial_k \zeta^m(t,\bx)+
\mG_{ijkm}(t,\bx)\partial_l \zeta^m(t,\bx)
 \ . \nonumber \\
\end{eqnarray}
It can be also shown that $K_{ij}$
  transform
covariantly under (\ref{fpd})
\begin{eqnarray}
K'_{ij}(t',\bx')=
K_{ij}(t,\bx)-K_{ik}(t,\bx)\partial_j
\zeta^k(t,\bx)
-\partial_i\zeta^k(t,\bx) K_{kj}(t,\bx)
\ .
\nonumber \\
\end{eqnarray}
Using these formulas it is easy to see
that the expression
$K_{ij}\mG^{ijkl}K_{kl}$ is invariant
under (\ref{fpd}).

Finally we consider the expression
$E^{ij}\mG_{ijkl}E^{kl}$. As was shown
in  \cite{Kluson:2009rk}  $E^{ij}$
transforms under (\ref{fpd}) as
\begin{equation}
E^{'ij}(t',\bx')=
E^{ij}(t,\bx)+\partial_k \zeta^i(t,\bx)
E^{kj}(t,\bx)+E^{ik}(t,\bx)\partial_k \zeta^j(t,\bx)
\ .
\end{equation}
 Then
 $E^{ij}\mG_{ijkl}E^{kl}$ is
clearly invariant under (\ref{fpd}).
Finally if we use the fact that
\begin{equation}
N'(t')\sqrt{g(\bx',t')} dt'd^D\bx'=N(t)
\sqrt{g(\bx,t)}dt d^D\bx
\end{equation}
under (\ref{fpd}) we find
 that the action
(\ref{LFR}) is invariant under
foliation  preserving diffeomorphism
(\ref{fpd}).

\section{The Second Version of $f(R)$
Ho\v{r}ava-Lifshitz  Theory of Gravity}
In this section we present the second
way how to define $f(R)$
Ho\v{r}ava-Lifshitz theories of
gravity. We begin with known $f(R)$
theory of gravity with the action
\begin{equation}\label{SR4}
S=-\frac{1}{\kappa^2}
\int dt d\bx N \sqrt{g}
f(R^{(4)})
\ ,
\end{equation}
where $R^{(4)}$ is four dimensional
covariant Ricci scalar. Then in order
to find $f(R)$ Ho\v{r}ava-Lifshitz
theory of gravity we simply  replace
$R^{(4)}$ in (\ref{SR4})
 with the combination
$ K_{ij}K^{ij}-\lambda^2
K-E^{ij}\mG_{ijkl}E^{kl}$ that appears
in the Ho\v{r}ava-Lifshitz theory that
obeys the detailed balance condition
\footnote{We would like to stress that
this procedure can be easily
generalized to theories that break the
detailed balance condition. Generally,
we could replace $R^{(4)}$ with
$K_{ij}K^{ij}-\lambda^2 K-\mathcal{V}$
where the explicit form of
$\mathcal{V}(g)$ was given in
\cite{Sotiriou:2009bx}.}.

As in previous section we introduce the
auxiliary fields $A$ and $B$ and write
the action (\ref{SR4}) in the
equivalent form
\begin{equation}\label{sFRa}
S=\kappa^2\int dt d^D\bx \sqrt{g}N
(B(K_{ij}K^{ij}-\lambda^2 K^2-E^{ij}
\mG_{ijkl}E^{kl}-A)+f(A)) \ .
\end{equation}
Clearly the solution of the  equation
of motion for $B$ reproduces the
original action while the equation of
motion for $A$ implies $f'(A)=B$. Inserting
this result into (\ref{sFRa}) we find
\begin{equation}
\label{sFRa2}
S=\kappa^2\int dt d^D\bx \sqrt{g}N
(f'(A)(K_{ij}K^{ij}-\lambda^2 K^2-E^{ij}
\mG_{ijkl}E^{kl}-A)+f(A)) \ .
\end{equation}
Our goal is to find the Hamiltonian
corresponding to the action
(\ref{sFRa}). Even if this action
is invariant under foliation preserving
diffeomorphism we can consider
more general case and presume that
the theory does not obey the projectability
condition so that $N$ depends on $\bx$.
Then we
introduce the set of conjugate
momenta $\pi^{ij}(\bx),
\pi^i(\bx),\pi_N(\bx)$ and
$P_A(\bx),P_B(\bx)$ where
\begin{eqnarray}
\pi^{ij}=\frac{\delta S}{\delta
\partial_t g_{ij}}=
\kappa^2 \sqrt{g}B \mG^{ijkl}K_{kl} \ ,
\nonumber \\
\end{eqnarray}
and where the remaining momenta form
the primary constraints of the theory
\begin{eqnarray}
 \pi_N(\bx)\approx 0 \ , \quad
\pi^i(\bx)\approx 0 \ , \quad  P_A(\bx)\approx 0
\ , \quad  P_B(\bx)\approx 0 \ .
\nonumber \\
\end{eqnarray}
Then it is straightforward  to find
corresponding  Hamiltonian
\begin{eqnarray}
H&=&\int d^D\bx \mH(\bx)= \int d^D\bx (N(\bx)
\mH_0(\bx)+N_i(\bx)\mH^i(\bx)+\nonumber \\
&+& v_i(\bx)\pi^i(\bx)+v_N(\bx)
\pi_N(\bx)
+v^A(\bx)P_A(\bx)+v_B(\bx)P^B(\bx)) \ , \nonumber \\
\mH_0&=& \left[\frac{1}{\kappa^2
B\sqrt{g}} \pi^{ij}\mG_{ijkl}\pi^{kl}+
\kappa^2 \sqrt{g}
B(E^{ij}\mG_{ijkl}E^{kl}+A)- \kappa^2
\sqrt{g}
f(A)\right] \ , \nonumber \\
\mH^i&=&-2\nabla_j\pi^{ij} \ . \nonumber \\
\end{eqnarray}
Since the constraints
$\pi^i(\bx)\approx 0 \ ,
\pi_N(\bx)\approx 0$ have to be preserved
during the time evolution of the
systems we find
\begin{eqnarray}
\partial_t \pi(\bx)&=& \pb{\pi_N(\bx),H}=  -\mH_0(\bx)\approx 0 \ , \nonumber \\
\partial_t \pi^i(\bx)&=&
\pb{\pi^i(\bx),H}=-\mH^i(\bx)\approx 0
\ .
\nonumber \\
\end{eqnarray}
On the other hand the requirement of
the preservation of the constraints
$P_A(\bx)\approx 0 \ , P_B(\bx)\approx 0$ during
the time evolution of the system imply
\begin{eqnarray}
\partial_t P_A(\bx)&=&
\pb{P_A(\bx),H}=
G_A(\bx)\approx 0
 \nonumber \\
\end{eqnarray}
and
\begin{eqnarray}
\partial_t P_B(\bx)&=&\pb{P_B(\bx),H}
=  N G_B(\bx) \approx 0 \ ,
\end{eqnarray}
where
\begin{eqnarray}
G_A&=&-B+f'(A) \ , \nonumber \\
G_B&=& \frac{1}{\kappa^2 B^2\sqrt{g}}
\pi^{ij}\mG_{ijkl}\pi^{kl}- \kappa^2
 \sqrt{g}(E^{ij}\mG_{ijkl}E^{kl}+A) \ .
 \nonumber \\
\end{eqnarray}
 Now we calculate the Poisson
brackets of these constraints when
we again introduce the notation
$\Phi_I=(P_A,G_A,P_B,G_B) \ ,
I=1,2,3,4$. Then we have
\begin{eqnarray}
\pb{\Phi_1(\bx),\Phi_2(\by)}&=& -f''(A)
(\bx)\delta(\bx-\by) \ ,  \nonumber
\\
\pb{\Phi_1(\bx),\Phi_3(\by)}&=&0 \ ,
 \nonumber \\
\pb{\Phi_1(\bx),\Phi_4(\by)}&=&
 \kappa^2
B(\bx)\sqrt{g}(\bx)\delta(\bx-\by) \ ,
\nonumber \\
\pb{\Phi_2(\bx),\Phi_3(\by)}&=&-\delta(\bx-\by)
\ ,
\nonumber \\
\pb{\Phi_2(\bx),\Phi_4(\by)}&=&0 \ , \nonumber \\
\pb{\Phi_3(\bx),\Phi_4(\by)}&=&
\kappa^2
\sqrt{g}(E^{ij}\mG_{ijkl}E^{kl}+A)
(\bx)\delta(\bx-\by) \ .
\nonumber \\
\end{eqnarray}
We see that $\Phi_I$ is collection of
the second class constraints that can
be set strongly to zero and explicitly
solved \footnote{It the similar way as
in previous section we can easily show
that the Dirac bracket
$\pb{g_{ij}(\bx),\pi^{kl}(\by)}_D$
coincides with the Poisson bracket
$\pb{g_{ij}(\bx),\pi^{kl}(\by)}$.}.
First of all, $\Phi_1=\Phi_3=0$ imply
that $P_A=P_B=0$. Further, from
$\Phi_2=0$ we  find
\begin{equation}\label{Bf'A}
B=f'(A) \ .
\end{equation}
Finally the constraint $\Phi_4=0$
together with (\ref{Bf'A})  gives
\begin{equation}
\frac{1}{\kappa^2 f'^2(A)\sqrt{g}}
\pi^{ij}\mG_{ijkl}\pi^{kl}=\kappa^2
\sqrt{g}(E^{ij}\mG_{ijkl}E^{kl}+A) \ .
\end{equation}
 Again, for known $f$
the equation above allows us to find
$A$ as a function of $
\frac{1}{\sqrt{g}}
\pi^{ij}\mG_{ijkl}\pi^{kl}$ and
$\sqrt{g}E^{ij}\mG_{ijkl}E^{kl}$ so
that
\begin{equation}
A=\Psi(\frac{1}{\sqrt{g}}
\pi^{ij}\mG_{ijkl}\pi^{kl},
\sqrt{g}E^{ij}\mG_{ijkl}E^{kl}) \ .
\end{equation}
Using these results we finally  find
the Hamiltonian constraint $\mH_0$ in
the form
\begin{eqnarray}\label{mH0nc}
\mH_0=
\frac{2}{\kappa^2\sqrt{g}f'(\Psi)}
\pi^{ij}\mG_{ijkl}\pi^{kl}
-\kappa^2 \sqrt{g}f(\Psi) \ .
\nonumber \\
\end{eqnarray}
\vskip .2in \noindent {\bf
Acknowledgements:} This work was
 supported by the Czech
Ministry of Education under Contract
No. MSM 0021622409.

\newpage

\end{document}